\documentclass[usenatbib,useAMS,preprint,psfig2]{mn2e}
\usepackage{times}
\usepackage{amsmath}
\usepackage{amssymb,amsfonts}
\usepackage{graphicx}
\usepackage{epstopdf}
\usepackage{verbatim}
\input psfig.sty

\def\gapprox{\lower.4ex\hbox{$\;\buildrel >\over{\scriptstyle\sim}\;$}}
\def\lapprox{\lower.4ex\hbox{$\;\buildrel <\over{\scriptstyle\sim}\;$}} \def\be{\begin{equation}}
\def\be{\begin{equation}}
\def\ee{\end{equation}}
\def\bea{\begin{eqnarray}}
\def\eea{\end{eqnarray}}
\catcode`\@=11
\font\tenmib=cmmib10
\font\tensyb=cmbsy10
\font\tenbi=cmmib10 
\newfam\bifam \def\bi{\fam\bifam\tenbi} 
\textfont\bifam=\tenbi  

\def\unboldmath{\everymath{}\everydisplay{}
          \textfont\@ne\teni 
          \textfont\tw@\tensy
          }
\def\boldmath{$\!\!$\relax\everymath{\mit}\everydisplay{\mit}
        \textfont\@ne\tenmib
        \textfont\tw@\tensyb 
        \relax}%
\catcode`\@=12
\def\bg#1{\hbox{\boldmath$#1$}}  

\begin{document}
\title[Parabolic arcs]{Interpretation of parabolic arcs in pulsar secondary spectra}
\author[Walker et al.]{M.A.~Walker$^1$, D.B.~Melrose$^1$, D.R.~Stinebring$^2$, C.M.~Zhang$^1$\\
1. School of Physics, University of Sydney, NSW 2006, Australia\\
2. Oberlin College, Department of Physics and Astronomy, Oberlin, OH 44074, U.S.A.}

\date{\today}

\maketitle

\begin{abstract}
Pulsar dynamic spectra sometimes show organised interference
patterns; these patterns have been shown to have power spectra
which often take the form of parabolic arcs, or sequences of
inverted parabolic arclets whose apexes themselves follow
a parabolic locus. Here we consider the interpretation of
these arc and arclet features. We give a statistical formulation
for the appearance of the power spectra, based on the stationary
phase approximation to the Fresnel-Kirchoff integral.
We present a simple analytic result for the power-spectrum
expected in the case of highly elongated images, and
a single-integral analytic formulation appropriate to the
case of axisymmetric images. Our results are illustrated
in both the ensemble-average and snapshot regimes.
Highly anisotropic scattering appears to be an important
ingredient in the formation of the observed  arclets.
\end{abstract}

\begin{keywords}
Scintillations --- pulsars --- interstellar medium
\end{keywords}

\section{Introduction}\label{sec:intro}
At low radio frequencies, diffraction from inhomogeneities in
the ionised component of the InterStellar Medium (ISM)
leads to multi-path propagation from source to observer.
Together with the refraction introduced by large-scale
inhomogeneities, this leads to a variety of observable
phenomena --- image broadening,
image wander and flux variations, for example, see Rickett (1990).
If the source is very small, as in the case of radio pulsars,
then the electric field has a high level of coherence
amongst the various propagation paths, and persistent,
highly visible interference fringes are seen in the
low frequency radio spectrum. These fringes evolve with time,
due to the motions of source, observer and ISM. Mostly
the observed fringes can be understood in terms of wave
propagation through a random medium, with the 
electron density inhomogeneities
being described by a power-law spectrum, close to the Kolmogorov
spectrum, and physically identified with a turbulent cascade
(Armstrong, Rickett and Spangler 1995; Cordes, Weisberg and
Boriakoff 1985; Lee and Jokipii 1976). However,
it has been known for many years that some pulsars, at
some epochs, also exhibit organised patterns in their
dynamic spectra -- drifting bands, criss-cross patterns
and periodic fringes, for example (Gupta, Rickett and Lyne 1994;
Wolsczan and Cordes 1987;
Hewish, Wolszczan and Graham 1985; Roberts and Ables 1982;
Ewing et al 1970). The interpretation of
these patterns has never been entirely clear, although
several authors have suggested an association with
enhanced refraction in the ISM (Rickett, Lyne and Gupta
1997; Hewish 1980; Shishov 1974).  

Recently a significant breakthrough in data analysis has
occurred, with the recognition that these organised patterns
are more readily apprehended by working with the so-called
``secondary spectrum'', i.e. the power spectrum of the dynamic
spectrum (Stinebring et al 2001): in this domain the very
complex patterns seen in the dynamic spectrum typically
appear as an excess of power along the locus of either
a single parabola, or a collection of inverted parabolic
arcs (Hill et al 2003; Stinebring et al 2004).
Previous investigations have demonstrated
why parabolic structures are generic features of the
secondary spectrum (Stinebring et al 2001; Harmon
and Coles 1983), but the properties of the observed
secondary spectra are not yet understood in detail.
In this paper we develop a statistical theory of the
power distribution in pulsar secondary spectra, based
on the stationary phase approximation, and present
some analytic results for simple forms of the scattered
image. We compare our results with existing data and
argue that highly anisotropic scattering is required
to form some of the observed secondary spectra. Our
results are consistent with, and complementary to, those
of Cordes et al (2003).

This paper is organised as follows. We begin by developing
a model secondary spectrum, based on the stationary phase
approximation, in \S2, leading to analytic models for
the ensemble-average secondary spectra of linear and
axisymmetric images (\S3), and Monte Carlo simulations
of snapshot secondary spectra (\S4). In \S5 we consider the
appearance of additional, off-centre image components; we
compare our models with existing data in \S6.

\section{The stationary phase approximation}
Consider a radio wave propagating over a distance
$D_p$ from a source to the observer, and passing
{\it en route\/} through a phase-changing screen
at a distance $D_s$ from the observer. We consider
only the case of a point-like source, implying a
large coherent patch;
this is a sensible approximation for pulsars, which
are known to be extremely compact sources. Let ${\bf x}$
denote the 2D position on the phase screen, then the
wave amplitude at a point ${\bf r}$ in the observer's
plane is
\begin{equation}u({\bf r})={-i\over
{2 \pi r_{\rm F}^{2}}} \int d^2{\bf x}\;\exp({i\Phi}),
\end{equation}
and to high accuracy the phase, $\Phi$, is given by
\begin{equation}
\Phi=\phi({\bf x})+{{({\bf x}-\beta{\bf r})^2}
 \over{2r_{\rm F}^2}},
\end{equation}
with ${r_{\rm F}^{2}}:=\beta\,D_s/k\,$, $\beta\equiv1-D_s/D_p$,
and $k=2\pi\nu/c$ for frequency $\nu$. Here the pulsar is
assumed to lie at the origin of the transverse coordinates
(${\bf r}$ and ${\bf x}$). In the trivial
case where $\phi=0$ everywhere this yields unit wave
amplitude at the observer --- the usual spherical
wave amplitude variation, $\exp(i\,kz)/z$, having been
absorbed into this normalisation.

For many purposes the exact integral formulation of eq. 1 can
be adquately represented by a sum over a finite number
of discrete points -- the stationary phase points
-- for which the derivative of the exponent with respect
to the variable of integration vanishes (see, e.g., Gwinn
et al 1998). This
replacement can be made because these points, and
their immediate surroundings, dominate the integral:
where the phase is not stationary, the integrand
oscillates, changing sign rapidly, and there is
little net contribution to the integral.
The condition for a point of stationary phase is
$\vec{\nabla}\Phi=0$, or
\begin{equation}
\vec{\nabla}\phi+{{\bf x}-\beta{\bf r}\over r_{\rm F}^{2}}=0.
\end{equation}
For each solution of equation 3, ${\bf x}={\bf x}_i({\bf r})$
with $i=1...N$, we expand the integrand to second order in ${\bf x}$
about the stationary point. It is 
assumed that these points are all well separated, so that this 
procedure may be applied to each point separately. The integral 
taken around the $i\,$th point gives a contribution
\begin{equation}
u_i({\bf r})=\sqrt{\mu_i}\,\exp({i\Phi_i}),
\end{equation}
where $\mu_i$ is the ``magnification'', and $\Phi_i$ the phase
for each path. The magnification is determined from the
phase curvature introduced by the screen:
\begin{equation}
\mu^{-1}=(1-\kappa)^2 - \gamma^2,
\end{equation} 
in terms of the convergence, $\kappa$,
and shear $\gamma:=\sqrt{\gamma_1^2+\gamma_2^2}$, with
\begin{equation}
\kappa=-{{r_F^2}\over{2}}\left[{{\partial^2\phi}\over{\partial x^2}}
+{{\partial^2\phi}\over{\partial y^2}}\right]
\end{equation} 
and
\begin{equation}
\gamma_1=-{{r_F^2}\over{2}}\left[{{\partial^2\phi}\over{\partial x^2}}
-{{\partial^2\phi}\over{\partial y^2}}\right],\qquad
\gamma_2=-r_F^2{{\partial^2\phi}\over{\partial x\partial y}},
\end{equation} 
where ${\bf x}=(x,y)$. The total electric field is then
\begin{equation}
u({\bf r})=\sum_i
{u_i({\bf r})}, 
\end{equation}
and the total intensity is
\begin{equation}
I(\nu,t)=u^*u\simeq\left|\sum_{i=1}^Nu_i\right|^2=\sum_{i,j=1}^N\sqrt{\mu_i\mu_j}
\cos\Phi_{ij},
\end{equation}
with $\Phi_{ij}\equiv\Phi_{i}-\Phi_{j}$. This result
is valid not only for the case of strong scattering, where
$N\gg1$ -- with each of the $N$ points coinciding with a
speckle in the image -- but also for lens-like (refractive)
behaviour where there may be only
a very small number of stationary phase points.

The result we have just derived is a description of
the dynamic spectrum of the source, because each of
the $\Phi_i$ is a function of frequency and time.
We will assume that the magnifications change only slowly in
comparison with the phases. In some circumstances
-- when the point under consideration is close to
a critical curve (corresponding to the source lying
close to a caustic) -- the $\mu_i$ are expected to change
rapidly with time and frequency; we will not consider
such cases, and henceforth we take the $\mu_i$ to
be constant. We can approximate the $\Phi$ variations
as linear in  frequency and time, over a total observing
time $\Delta t$, centred on $t_0$, and bandwidth
$\Delta\nu$ centred on $\nu_0$:
\begin{equation}
\Phi_{ij}\simeq\Phi_{ij}^0+
{{\partial\Phi_{ij}}\over{\partial t}}(t-t_0)+
{{\partial\Phi_{ij}}\over{\partial\nu}}(\nu-\nu_0),
\end{equation}
where $\Phi_{ij}^0\equiv\Phi_{ij}(\nu_0, t_0)$.
Taking the Fourier Transform of equation 9 then yields
$\tilde{I}:=\tilde{I}(f_\nu,f_t)$:
\begin{eqnarray}
\tilde{I}={{\Delta t\,\Delta\nu}\over{4\pi}}
\sum_{i,j=1}^N\sqrt{\mu_i\mu_j}\bigl\{\exp[i\Phi_{ij}^0]\;
{\rm sinc}[\pi\,\Delta t\,(f_t+F_{t,ij})]\cr
\qquad\times\;\;\; {\rm sinc}[\pi\,\Delta \nu\,(f_\nu+F_{\nu,ij})]
\;\,\qquad+\qquad\exp[-i\Phi_{ij}^0]\cr
\qquad\times\;\;\;{\rm sinc}[\pi\,\Delta t\,(f_t-F_{t,ij})]
\;{\rm sinc}[\pi\,\Delta \nu\,(f_\nu-F_{\nu,ij})]\bigr\}.
\end{eqnarray}
For large $\Delta t,\,\Delta\nu$ this result can be approximated by
\begin{eqnarray}
\tilde{I}\simeq{{1}\over{4\pi}}
\sum_{i,j=1}^N\sqrt{\mu_i\mu_j}
\bigl\{\exp[i\Phi_{ij}^0]\;\delta(f_t+F_{t,ij})\;\delta(f_\nu+F_{\nu,ij})\cr
+\quad\exp[-i\Phi_{ij}^0]\;\delta(f_t-F_{t,ij})\;\delta(f_\nu-F_{\nu,ij})\bigr\},
\end{eqnarray}
where $\delta$ denotes the Dirac Delta Function, and
\begin{equation}
F_{t,ij}\equiv{1\over{2\pi}}{{\partial\Phi_{ij}}\over{\partial t}},
\qquad
F_{\nu,ij}\equiv{1\over{2\pi}}{{\partial\Phi_{ij}}\over{\partial \nu}}.
\end{equation}
Our model ``secondary spectrum'' is therefore (the power
spectrum of equation 9):
\begin{eqnarray}
P(f_\nu,f_t)=\tilde{I}^*\tilde{I}
\simeq{ {\Delta t\,\Delta\nu}\over{(4\pi)^2}}\sum_{i,j=1}^N\mu_i\mu_j
\big[\delta\left(f_t+F_{t,ij}\right)\cr
\qquad\times\delta\left(f_\nu+F_{\nu,ij}\right) +
\delta\left(f_t-F_{t,ij}\right)\;
\delta\left(f_\nu-F_{\nu,ij}\right) \big].
\end{eqnarray}

The foregoing analysis shows that
the various interference terms contribute
power to the secondary spectrum in proportion to $\mu_i\mu_j$,
at specific fringe frequencies given by the derivatives
of the phase differences, $\Phi_{ij}$. We note that the
secondary spectrum is symmetric under the operation
$(f_\nu,f_t)\rightarrow(-f_\nu,-f_t)$. This arises because
$I(\nu,t)$ is a real quantity, reflecting the interchange
asymmetry $\Phi_{ij}=-\Phi_{ji}$. Another manifestation
of the symmetry is the fact that fringes with wavevector
$(f_\nu,f_t)$, and those with wavevector $(-f_\nu,-f_t)$
are indistinguishable in the dynamic spectrum. Because of
the symmetry in $P$, we will henceforth consider only the
half-plane $f_\nu\ge0$, for which the secondary spectrum
just derived can be rewritten as
\begin{equation}
P(f_\nu,f_t)\propto\sum_{i,j=1}^N\mu_i\mu_j\,
\delta\left(f_t-F_{t,ij}\right)\;\delta\left(f_\nu-F_{\nu,ij}\right),
\end{equation}
and this is the form we shall make use of subsequently.

\subsection{Phase relationships in multipath propagation}
The trajectory associated with the $i\,$th stationary phase
point exhibits a phase $\Phi_i=\Phi({\bf x}_i, {\bf r}, \nu, t)$,
given by equation 2,
relative to the direct line-of-sight to the source in the
absence of a screen. The derivatives of $\Phi$ with respect
to these four variables can be evaluated as follows. Quite
generally, differentiating equation 2 with respect to ${\bf r}$ yields
\begin{equation}
\left({{\partial\Phi}\over{\partial {\bf r}}}\right)_{{\bf x},\nu,t}
= -{\beta \over{r_F^2}}({\bf x} - \beta{\bf r}).
\end{equation}
Similarly, differentiating equation 2 with respect to $\nu$ gives
\begin{equation}
\left({{\partial\Phi}\over{\partial {\nu}}}\right)_{{\bf x},t,{\bf r}}
={1\over\nu}\left[{{({\bf x} - \beta{\bf r})^2}\over{2r_F^2}}
- 2\phi \right],
\end{equation}
where we have made use of the result
\begin{equation}
\left({{\partial\,\log\,\phi}\over{\partial\,\log\,\nu}}\right)_t=-2,
\end{equation}
which is appropriate for radio-wave propagation in the Galaxy
where the refractive index is dominated by the free-electron
contribution. Differentiating equation 2 with respect to ${\bf x}$ gives
\begin{equation}
\left({{\partial\Phi}\over{\partial {\bf x}}}\right)_{\nu,t,{\bf r}}=
\vec{\nabla}\phi+{{\bf x}-\beta{\bf r}\over r_{\rm F}^{2}}=0,
\end{equation}
where the final equality (equation 3) applies for stationary phase points.
Finally, differentiating equation 2 with respect to $t$ yields
\begin{equation}
\left({{\partial\Phi}\over{\partial {t}}}\right)_{{\bf x},{\bf r},\nu}=
\left({{\partial\phi}\over{\partial {t}}}\right)_{{\bf x},\nu}.
\end{equation}
We will assume that the phase structure in the screen is ``frozen'',
and thus is convected with the screen velocity, ${\bf v}_s^\prime$, so that
\begin{equation}
{\bf v}_s^\prime\cdot\vec{\nabla}\phi +
{{\partial\phi}\over{\partial t}}=0
\end{equation}
for all paths. Using the stationary phase condition then gives us
\begin{equation}
\left({{\partial\Phi_i}\over{\partial {t}}}\right)_{{\bf x},{\bf r},\nu}=
{1\over{r_F^2}}({\bf x} - \beta{\bf r})\cdot{\bf v}_s^\prime.
\end{equation}

Making use of equations 16 and 22, it is a straightforward exercise
to evaluate the temporal fringe frequency for a given stationary
phase path:
\begin{eqnarray}
2\pi\,F_{t,i}=\left({{\partial\Phi_{i}}\over{\partial t}}\right)_{\nu}
=\left({{\partial\Phi_i}\over{\partial {t}}}\right)_{{\bf x},{\bf r},\nu}
+{\bf v}_o^\prime\cdot
\left({{\partial\Phi}\over{\partial {\bf r}}}\right)_{{\bf x},\nu,t}\cr
= {1\over{r_F^2}}({\bf x} - \beta{\bf r})\cdot
({\bf v}_s^\prime - \beta{\bf v}_o^\prime),
\end{eqnarray}
where the observer's velocity is ${\bf v}_o^\prime$.
If the source velocity is ${\bf v}_p$, relative to some
reference frame, then the
actual space velocities of the screen and observer
are ${\bf v}_{s,o}={\bf v}_{s,o}^\prime+{\bf v}_p$.
Only the components of these velocities perpendicular
to the line-of-sight are of relevance here. Defining
${\bf v}_\perp\equiv{\bf v}_s-\beta{\bf v}_o-(1-\beta){\bf v}_p$,
where all velocities are to be understood as being
transverse components, allows us to write (dropping
the subscript $i$ which denotes the stationary phase
path under consideration)
\begin{equation}
F_t={1\over{\lambda\beta}} {\bg\theta}\cdot{\bf v}_\perp,
\end{equation}
where $\lambda=2\pi/k$, and ${\bg\theta}=({\bf x}-\beta{\bf r})/D_s$
is the angular separation between the path under
consideration and the direct line-of-sight to
the source.

Implicit in this result for the temporal
fringe frequency are the Doppler shifts due to the
motions of source, screen and observer, each of which
contributes to ${\bf v}_\perp$.
All of these terms are linear in the relevant velocity
and in the angular separation of the paths under
consideration, so for speeds $\sim300\,{\rm km\,s^{-1}}$,
typical of pulsars, and angles of order a couple of
milli-arcseconds, the fractional frequency shifts
are $\sim\theta v/c\sim10^{-11}$, hence beat periods
of order two minutes at a radio frequency of 1~GHz.
We note that these frequency shifts have not been
calculated from a relativistic formulation,
and are therefore not correct to second order in $v/c$.

The spectral fringe rate for stationary phase paths,
\begin{equation}
F_{\nu,i}\equiv{1\over{2\pi}}\left({{\partial\Phi_{i}}\over{\partial\nu}}\right)_t=
{1\over{2\pi}}\left({{\partial\Phi}\over{\partial {\nu}}}\right)_{{\bf x},t,{\bf r}},
\end{equation}
follows directly from equation 17.
Again dropping the subscript which denotes the particular
stationary phase path under consideration, we have
\begin{equation}
F_\nu={{D_s\theta^2}\over{2c\beta}} - {\phi\over{\pi\nu}}.
\end{equation}
The physical interpretation of the various terms 
contributing to $F_\nu$ is straightforward.
The first term in equation 26 is the geometric
contribution to the delay along the path under
consideration, while the second term is the delay
due to the propagation speed of the radio-wave
within the phase screen;
the former is frequency independent, while the latter
is dispersive. The geometric delay is expected to be
$\simeq\theta^2 D/c\sim10^{-5}{\rm s}$, for refraction
angles $\theta\sim2$~milli-arcseconds, and source
distance $\sim1$~kpc.

Equations 24 and 26 apply to the individual stationary
phase paths, from which we can compute $F_{t,ij}, F_{\nu,ij}$,
via $\Phi_{ij}\equiv\Phi_i-\Phi_j$.
Having identified the physical interpretation of the
terms contributing to $F_t,\,F_\nu$, as being Doppler
shifts and delays, respectively, we see that the
secondary spectrum can be regarded as a
``delay-Doppler diagram'', with the power appearing
at locations corresponding to differences in
delay/Doppler-shift between the various interfering paths
(Harmon and Coles 1983; Cordes et al 2004).

\subsection{Neglect of dispersive delays}
From equations 24 and 26 it can be seen that if the
dispersive delay (the last term in eq. 26) is
neglected, then the fringe frequencies obey a quadratic
relationship because $F_t$ depends linearly on
${\bg\theta}$, whereas $F_\nu$ varies
quadratically. Neglect of the dispersive delay is a
good approximation, for stationary phase paths, if
\begin{equation}
\left|({\bf x - \beta r})\cdot{{\vec{\nabla}\phi}\over{\phi}}\right|\gg4.
\end{equation}
In other words the screen phase, $\phi$, must be changing
on a length-scale ($L$) which is small compared with the distance
of the path from the optical axis (the direct line-of-sight to the
source).  The meaning of this result can be seen most
easily if we consider two limiting cases: purely scattering
(diffractive), and purely lens-like (refractive) phase screens.
In the pure scattering case, ${\vec{\nabla}\phi}$ changes sign
on a length scale $\sim L\ll r_F$ much smaller than
the typical separation $\left|{\bf x - \beta r}\right|\ga r_F$.
Consequently the neglect of
dispersive delays is a natural approximation to make
in the case of a pure scattering screen. However, the
converse is not necessarily true; to see this we can consider a
screen in which the phase increases as $({\bf x - \beta r})^n$.
In this case the condition given in equation 27 becomes $n\gg4$,
which clearly can be satisfied even though there is no
small-scale phase structure (no scattering) in the phase screen. 

In the major part of the present paper we
will confine ourselves to consideration of the
case where the dispersive delays are negligible,
returning to the issue only in \S5.
This restriction is motivated in large part by
the attendant simplification of results, and encouraged
by the fact that the resulting model seems to offer
a fair description of many of the observed phenomena.
Noting that the fringe frequencies $F_{t,ij}$ and $F_{\nu,ij}$
include common normalising factors (eqs 24 and 26), it
is useful to introduce the following coordinates
(Stinebring et al 2001;  Hill et al 2003):
\begin{equation}
q:={{\beta\lambda}\over{|{\bf v}_\perp|}}f_t=\theta_{1x}-\theta_{2x},
\end{equation}
and
\begin{equation}
p:={{2c\beta}\over{D_s}}f_\nu={\bg\theta}_1^2-{\bg\theta}_2^2.
\end{equation}
For simplicity we have chosen the $x$-axis to lie
along ${\bf v}_\perp$. 

\section{Statistical model of the secondary spectrum}
Distant sources ($D\ga1$~kpc) observed at low radio
frequencies are expected to be in the regime of
strong scattering, for which there are many paths
of stationary phase from source to observer. In
this circumstance it is appropriate to employ a
statistical treatment of the power distribution in
the secondary spectrum. In this section we present
such a treatment, with emphasis on the analytic
results which can be obtained for some specific
cases. These results should prove to be useful
for observers undertaking quantitative analysis
of the power distribution in secondary spectra.

The intensity distribution in the scattered image
can be written in terms of a probability distribution,
$g({\bg\theta})$, for the power in the stationary
phase paths. Equation 15 shows that the various
interference terms contribute power to the secondary
spectrum in proportion to $\mu_i\mu_j$, and the
expectation value of this quantity
is $g({\bg\theta}_1)g({\bg\theta}_2)\,d^2{\bg\theta}_1
d^2{\bg\theta}_2$, where we have replaced the
specific labels $i,j$ by the generic pair label $1,2$.
Including only interference terms
which contribute to the particular fringe frequencies
$p,q$ considered, the expected power distribution in
the secondary spectrum can therefore be described by
\begin{equation}
P=\int\!d^2{\bg\theta}_1d^2{\bg\theta}_2
\,g({\bg\theta}_1)g({\bg\theta}_2)\,
\delta(p-{\bg\theta}_1^2+{\bg\theta}_2^2)\delta
(q-\theta_{1x}+\theta_{2x}).
\end{equation}
Strictly this formulation should be interpreted as
yielding the ensemble-average (i.e. long-term
time-average) of $P$. However, in order to reach this
regime observationally one must average data over
time-scales which are long in comparison to the
refractive time-scale, and this is generally not
the case in practice. In principle one could average
the secondary spectra for a given target observed
over a period of several months or years, much longer
than the refractive time-scale, but in practice
this is not a sensible procedure because the large-scale
distribution of power in the secondary spectrum can
undergo profound changes on time-scales of weeks,
indicating that the statistical properties are not stationary
on these time-scales, so that the ensemble-average
regime cannot be reached. Instead we are obliged to
deal with data in the limit of almost instantaneous
sampling, or else in the average-image regime with
only a short averaging time. The various regimes are
characterised by an averaging time, $t$, which
stands in the following relationships to the
diffractive time-scale, $t_d$, and the refractive
time-scale, $t_r$ (Goodman and Narayan 1989):
$t\gg t_r$, ensemble average; $t_r\ga t>t_d$,
average; $t<t_d$, snapshot. For pulsars observed
at frequencies $\nu\sim400-1000$~MHz, representative
time-scales are $t_d\sim$~minutes, $t_r\sim$~weeks;
a single observation might last for $t\sim30$~minutes,
and is therefore generally in the average-image regime.
Some data may be in the snapshot regime, and this
regime is considered in \S4.
The expected image-plane power distributions in the
average and ensemble-average regimes are identical,
and thus the formulation given in equation 30 is
appropriate for much of the data. However, when
comparing our models with data we must bear in mind
that some deviations are expected simply because the
data are not ensemble averages.

\subsection{Image-plane probability distributions}
In order to compute the secondary spectrum via
equation 30, we need to specify the probability
distribution $g({\bg\theta})$.
The expected angular distribution of power is given
by the spatial Fourier Transform of the mutual coherence
function (Lee and Jokipii 1975):
\begin{equation}
 g({\bg\theta})\propto\int d^2{\bg\rho}\,
 \exp({-ik{\bg\rho}\cdot{\bg\theta}})\,\Gamma_2({\bg\rho}).
\end{equation}
For scattering by turbulence in a single screen, the
mutual coherence function is
\begin{equation}
\Gamma_2({\bg\rho})=\exp({-D({\bg\rho})}),
\end{equation}
where the phase structure function is given by
\begin{equation}
D({\bg\rho})=\langle[\phi({\bf x}+{\bg\rho})-\phi({\bf x})]^2
\rangle.
\end{equation}
In the present paper we will consider only structure functions
which are power-law in the separation ${\bg\rho}$, and
we parameterise these forms in terms of the field
coherence scale, $s_o$: $D(s_o)=1$. A characteristic
angular scale for the image is then $\theta_o=1/(k\,s_o)$.

\begin{figure}
\includegraphics[width=8cm]{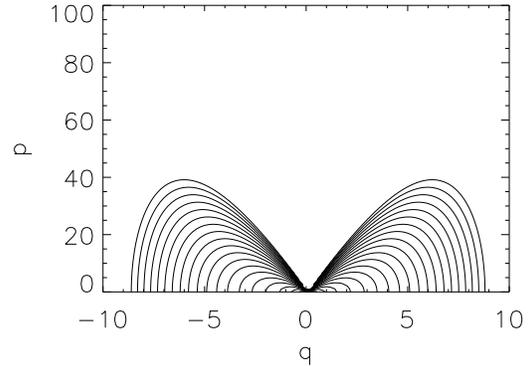}
\caption{``Secondary spectrum'' (or ``delay-Doppler''
diagram) for a linear image, at an angle $\psi=0$ to
the $x$-axis (defined by the effective transverse
velocity vector), with a Gaussian intensity profile
(defined by equation 39).
The axes are expressed in units of $\theta_o=1/k\,s_o$,
for $q$, and $\theta_o^2$ for $p$. Contour levels are
set at intervals of $3$~dB, and the lowest (outermost)
contour is at $-60$~dB relative to $P(0.1,0.1)$.}
\end{figure}
\begin{figure}
\includegraphics[width=8cm]{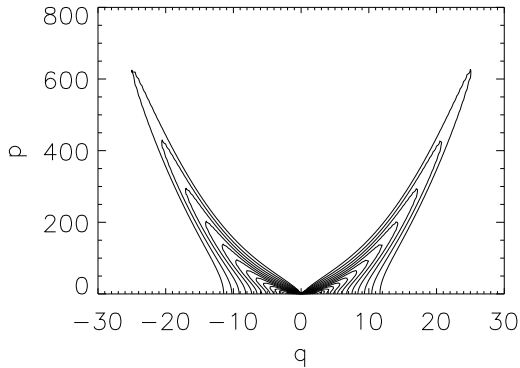}
\caption{Secondary spectrum for a linear image, as
figure 1 but for the image intensity profile given
by equation 41 (representing the case of scattering
by highly anisotropic Kolmogorov turbulence). Note
the difference in the scale of the axes relative
to figure 1; the contour levels are identical to those
in figure 1.}
\end{figure}

\subsection{Highly anisotropic images}
A useful model for highly anisotropic
images follows when one considers the limiting case of
a linear image.
A linear source making an angle $\psi$ with respect
to the $x$-axis corresponds to an image with
\begin{equation}
g({\bg\theta})=h(\theta_x\cos\psi+\theta_y\sin\psi)\,
\delta(-\theta_x\sin\psi+\theta_y\cos\psi),
\end{equation}
where the function $h$ denotes the variation in
intensity along the line. On inserting eq. 34 into eq. 30,
there are four integrals and four $\delta$-functions. The
probability $P(p,q)$ is then found by solving the four
simultaneous equations, implied by the four $\delta$-functions,
for $\theta_{\pm}=\theta_{x\pm}\cos\psi+\theta_{y\pm}\sin\psi$
and identifying $P(p,q)$ as $h(\theta_{+})h(\theta_{-})$
divided by the Jacobian of the transformation
to the variables specified by the
$\delta$-functions. The four simultaneous equations give
\begin{eqnarray}
\qquad\theta_{x+}={p\cos^2\psi+q^2\over2q},
\quad
\theta_{y+}=\tan\psi\,\theta_{x+},\hfill\cr
\qquad\theta_{x-}={p\cos^2\psi-q^2\over2q},
\quad
\theta_{y-}=\tan\psi\,\theta_{x-},
\end{eqnarray}
and hence
\begin{equation}\theta_{\pm}={p\cos^2\psi\pm q^2\over2q\cos\psi}.
\end{equation}
The Jacobian is $|q|$, hence one finds
\begin{equation}P(p,q)={h(\theta_{+})\,h(\theta_{-})\over|q|},
\end{equation}
with $\theta_{\pm}$ given by eq. 36.

\subsubsection{Specific examples of anisotropic images}
Here we take the general formulation given above
and apply it to two specific cases.
For simplicity we calculate only the case $\psi=0$,
corresponding to an image which is elongated
along the same direction as the effective velocity
${\bf v}_\perp$.

\begin{figure}
\includegraphics[width=8cm]{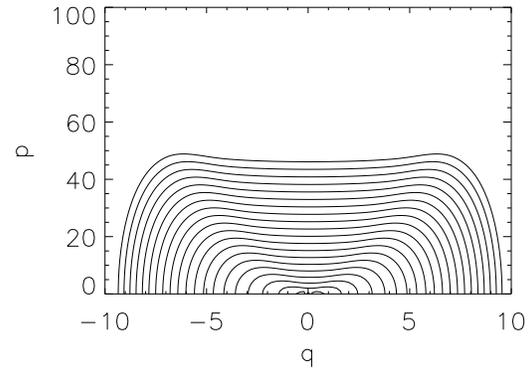}
\caption{``Secondary spectrum'' (or ``delay-Doppler''
diagram) for an axisymmetric image with a Gaussian
intensity profile (defined by equation 47).
The axes are expressed in units of $\theta_o=1/ks_o$,
for $q$, and $\theta_o^2$ for $p$. Contour levels are
set at intervals of $3$~dB, and the lowest (outermost)
contour is at $-60$~dB relative to $P(0.1,0.1)$.}
\end{figure}
\begin{figure}
\includegraphics[width=8cm]{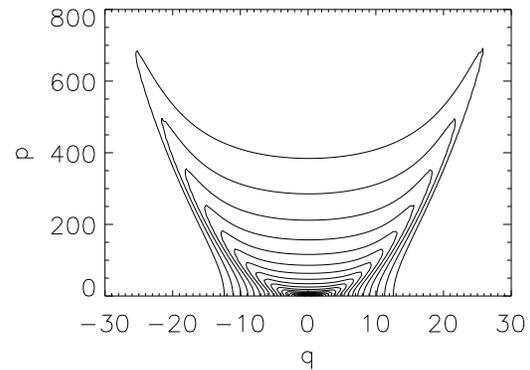}
\caption{Secondary spectrum for an axisymmetric image, as
figure 3 but for the image intensity profile given
by equation 48 (representing the case of scattering
by isotropic Kolmogorov turbulence). Note the difference
in the scale of the axes relative to figure 3; the contour
levels are identical to those in figure 3.}
\end{figure}

In the case of a quadratic phase structure function,
\begin{equation}
D({\bg\rho})=\left({{\rho_x}\over{s_o}}\right)^2,
\end{equation}
with no dependence on $\rho_y$, the resulting
image is Gaussian, with
\begin{equation}
h(\theta)={1\over{\sqrt{2\pi}\theta_o}}\exp(-\theta^2/4\theta_o^2).
\end{equation}
The secondary spectrum corresponding to this image
profile is shown in figure 1, with $q$ measured in
units of $\theta_o$, and $p$ in units of $\theta_o^2$.
The symmetry of the secondary spectrum is due to the
symmetry of the image around $\theta_x=0$. The contours
in figure 1 are spaced at intervals of 3~dB, with the
lowest (outermost) contour corresponding to $-60$~dB
relative to the peak --- the features shown here are
very faint. In this, and subsequent figures,
we have taken the peak value of the secondary spectrum
to be $P(0.1,0.1)$; normalising to these coordinates
is somewhat arbitrary, but normalising at $(0,0)$ is
not sensible because $P\to\infty$ there. The fact
that $P$ peaks for small $p,q$ simply reflects the
fact that the sum of the self-interference terms is
larger than the sum of cross-power terms in equation
15. The most striking feature of figure 1 is the absence
of power in the region $q=0,\,p>0$; this is a
consequence of the fact that we are considering a
linear image. The case $q=0$ corresponds to interference
between points which have the same value of $\theta_x$,
but we are considering an image where all of the power
is concentrated along a line with $\theta_y\propto\theta_x$,
so the secondary spectrum is devoid of cross-power at
small $q$ --- there is only the self-interference
which appears at $p=0$.

A particularly interesting case to consider is that
where the structure in the phase screen arises from
highly anisotropic Kolmogorov turbulence, in which case
\begin{equation}
D({\bg\rho})=\left({{\rho_x}\over{s_o}}\right)^{5/3}
\end{equation}
(Narayan and Hubbard 1988),
and the image profile cannot be expressed
in a simple analytic form. Numerically we find that
in this case the intensity profile assumes a power-law
form at large angles, with a power-law index of $-8/3$,
and we adopt the approximate form
\begin{equation}
h(\theta)\simeq {{h(0)}\over{1+(\theta/\theta_o)^{8/3}}},
\end{equation}
where $h(0)\equiv4\sin(3\pi/8)/{3\pi}$. This image
profile yields much more power scattered to large
angles than the Gaussian profile considered above.
The corresponding secondary spectrum is displayed
in figure 2; note the large increase in the area
plotted, relative to figure 1, even though
the contour levels are identical in the two
figures. This feature is a direct consequence of the
fact that the Kolmogorov phase screen yields an
image with much more power at large angles than
the Gaussian profile, yielding much larger delays
(i.e. much larger $p$) at a given contour level.
Figure 2 displays the same absence of power in
the region $p>0,\,q=0$ as seen in figure 1,
and for the same reasons. A new feature in figure
2, however, is the parabolic form $p=q^2$ delineated by
the contouring at $p,q\gg1$. This can be understood
by recognising that the dominant contributions to
secondary spectrum are due to interference
terms which include the highest intensity region of the
image, namely $|\theta|\ll\theta_o$ in this case. Referring
to equation 36, we see that this corresponds to
$|\theta_-|\ll\theta_o$, hence $|p-q^2|\la2|q|$. More generally,
for a linear image making an angle $\psi$ with the
$x$-axis, at large values of $p$ the secondary spectrum
will exhibit contours which cluster around $p=q^2\sec^2\psi$.

\subsection{Axisymmetric images}
{\it A priori} perhaps the most plausible model is one
in which the probability  does not depend on direction:
$g({\bg\theta})=g({\theta})$, i.e. the axisymmetric case. 
In this case two of the integrals in eq. 30
can be performed over the $\delta$-functions,
and a useful result requires that one perform a third integral
analytically. This would then leave a single integral that can
be evaluated either numerically, or with the use of simple
analytic approximations. In view of
the symmetry, the choice of polar coordinates is the most appropriate.
With this choice eq. 30 becomes
\begin{eqnarray}
\hfilneg P(p,q)=\int\! d\theta_1\theta_1d\psi_1\int\! d\theta_2\theta_2d\psi_2
\;g(\theta_1)\,g(\theta_2)\qquad\qquad\hfill\cr
\times\,\delta(p-\theta_1^2+\theta_2^2)\;\delta(q-\theta_1\cos\psi_1+\theta_2\cos\psi_2).
\end{eqnarray}
The integrals are carried out in the Appendix; the result is
\begin{eqnarray}
P(p,q)=2\int_0^\Theta
{d\theta_2\,\theta_2\,g(\theta_2)g\big(\sqrt{p+\theta_2^2}\big)
\over
[(\sqrt{p+\theta_2^2}+\theta_2)^2-q^2]^{1/2}}\qquad\qquad\hfill\cr
\qquad\qquad\times{\bi K}\left(\left[
{4\sqrt{p+\theta_2^2}\,\theta_2\over
(\sqrt{p+\theta_2^2}+\theta_2)^2-q^2}
\right]^{1/2}
\right)
\hfill\cr
\qquad\qquad
+\int_{|(p-q^2)/2q|}^\infty {d\theta_2\,\theta_2\,g(\theta_2)
g\big(\sqrt{p+\theta_2^2}\big)\over
[\sqrt{p+\theta_2^2}\,\theta_2]^{1/2}}\qquad\qquad\hfill\cr
\qquad\qquad\times{\bi K}\left(
\left[{(\sqrt{p+\theta_2^2}+\theta_2)^2-q^2\over
4\sqrt{p+\theta_2^2}\,\theta_2}
\right]^{1/2}
\right),
\end{eqnarray}
where ${\bi K}$ is the Complete Elliptic Integral,
and the upper limit of integration is
$\Theta=(p-q^2)/4|q| + |(p-q^2)/4q|$.
This result is not amenable to further analytic manipulation,
in general, and is our final form for general axisymmetric
images. In some cases it is possible to approximate
equation 43 with a simpler analytic result. For example:
if the probability distribution, $g(\theta)$, is very
sharply peaked at $\theta=0$, then for $p,q>0$ and
$p>q^2$ the second integral in eq. 43 can be neglected
(see equations A.11 and A.12), and we can make the
approximation ${\bi K}\simeq\pi/2$, so that
\begin{equation}
P(p,q)\simeq2\int_0^\infty{{{\rm d}\theta_2\;\theta_2\,g(\theta_2)
\,g(\sqrt{p})}\over{\sqrt{p-q^2}}}{\pi\over2}={1\over2}{{g(\sqrt{p})}\over{\sqrt{p-q^2}}}.
\end{equation}
However, in general the model secondary spectrum for
any given axisymmetric image must be evaluated numerically
from the formulation given in equation 43

\subsubsection{Specific examples of axisymmetric images}
For the general case of an isotropic, power-law structure
function:
\begin{equation}
D({\bg\rho})=\left({{\rho}\over{s_o}}\right)^\alpha,
\end{equation}
we have
\begin{eqnarray}
\qquad\qquad g({\bg\theta})\propto
\int d^2{\bg\rho}\,\exp[{-ik{\bg\rho}\cdot{\bg\theta}-D({\bg\rho})}],\cr
\qquad\qquad\qquad\propto\int_0^\infty dt\,t\,J_0(\theta t/\theta_o)\,
\exp({-t^\alpha}).
\end{eqnarray}
For $\alpha=2$ it is straightforward to show that the
resulting image profile is Gaussian, with
\begin{equation}
 g({\bg\theta})={1\over{4\pi\theta_o^2}}
 \exp(-\theta^2/4\theta_o^2).
\end{equation}
The secondary spectrum corresponding to this image is
shown in figure 3; comparing this to figure 1 (the
result for a linear, Gaussian image) we see that the
structure is broadly similar in the region $p\la q^2$,
but the axisymmetric image exhibits much more power
in the region $p>q^2$. This can be understood in the
following way: whereas
the linear images considered in \S3.2.1 possess a unique
relationship of the form $\theta_y\propto\theta_x$, so
that a small value of $q$ necessarily corresponds to
a small value of $p$, no such relationship exists for
the axisymmetric image --- even for $q=0$ (so $\theta_{x1}
=\theta_{x2}$), there are many contributing values of
$\theta_{y1},\,\theta_{y2}$, and hence there is power spread
over a range of values of $p$.

By contrast to the case $\alpha=2$, for isotropic Kolmogorov
turbulence, with $\alpha=5/3$, the integral does not
yield a simple analytic form. As with the highly anisotropic
case in \S3.2.1, evaluating numerically reveals a power-law
variation in the image surface-brightness, with
$g(\theta)\propto\theta^{-11/3}$ for $\theta\gg\theta_o$
--- there is much more power scattered to large angles
in the case of a Kolmogorov spectrum of turbulence
than for the case of a quadratic phase structure function.
We will use the following approximation for the case of
isotropic Kolmogorov turbulence:
\begin{equation}
 g({\bg\theta})\simeq {{g(0)}\over{1+(\theta/\theta_o)^{11/3}}},
\end{equation}
where $g(0)\equiv11\sin(6\pi/11)/{6\pi^2}$.
The corresponding secondary spectrum is shown in
figure 4; note the much expanded scale on both axes,
relative to figure 3, and the enhanced power around
the parabolic arc defined by $p=q^2$. Comparing to the
case of highly anisotropic
Kolmogorov turbulence, shown in figure 2, we find that
the axisymmetric case is similar to the anisotropic case
for $p\la q^2$, but exhibits much more power at $p>q^2$.
This same feature was found in the comparison between
linear and axisymmetric Gaussian images (see above),
and occurs for the same reason. 

\begin{figure}
\includegraphics[width=8cm]{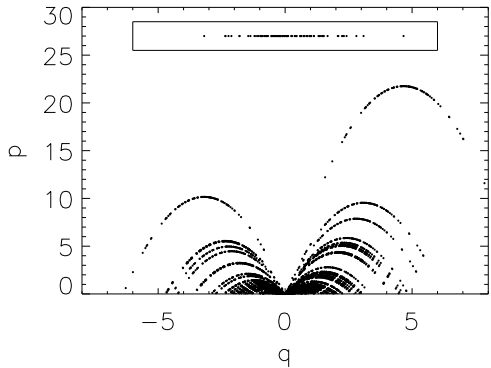}
\caption{Secondary spectrum (delay-Doppler diagram)
for a linear Gaussian image in the speckle regime (with
$\psi=0$, so the image is elongated along the direction
of ${\bf v}_\perp$). As
for the earlier figures, the axes here are given in
units of $\theta_o$ and $\theta_o^2$, for $q$ and
$p$, respectively. The expected image profile is
given by equation 39, and the actual image profile
is shown in the inset at the top of this figure.
The ensemble-average secondary spectrum for this
circumstance is shown in figure 1.}
\end{figure}
\begin{figure}
\includegraphics[width=8cm]{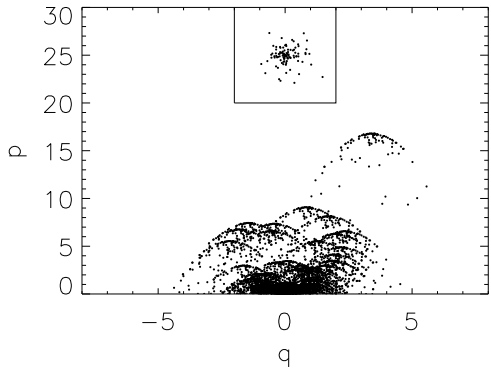}
\caption{As figure 5, but for the case of an axisymmetric
Gaussian image (as per equation 47) in the speckle regime;
the actual speckle image is shown in the inset box.
The ensemble-average secondary spectrum for this
circumstance is shown in figure 3.}
\end{figure}

\section{Snapshot secondary spectra}
The results described in \S3 are expectation values
for the secondary spectrum in various circumstances;
they represent the ensemble average -- i.e. the
average over many realisations of individual phase
screens -- and are thus appropriate to the long-term
time-averaged value of observations of secondary spectra.
As discussed in \S3, individual observations of a real
source are not expected to be in this regime. Usually
the results which are reported lie in the average-image
regime -- for which the model developed in \S3 is still
appropriate, albeit with the caveats given there -- but in
some instances the
data may be in the snapshot regime. In this case there
is only a single realisation of the randomness in the
phase screen contributing to the observed secondary
spectrum, and the phase screen can be represented by a
particular set of
stationary phase points whose specific locations are
unknown. Obviously the snapshot secondary spectra cannot
be predicted in detail, but it is important to understand
their general appearance. We give two examples of snapshot
secondary spectra, for the cases of linear and
axisymmetric Gaussian images; these results may be compared
directly with those of \S3.2.1 and \S3.3.1, respectively.

\subsection{Monte Carlo realisations of snapshot spectra}
To demonstrate the appearance of a secondary spectrum
from a single realisation of a phase screen, rather than
the long-term average value, we return to the formulation
given in \S2. To generate a secondary spectrum from 
equation 15 we need to specify the locations and magnifications
for each of the stationary phase points contributing
to the image. If the expected image profile is specified,
e.g. a Gaussian profile in the case of a quadratic phase
structure function, then we can generate an individual
realisation, consistent with the expected properties,
by the Monte Carlo method. The expected magnification
of a stationary phase point is independent of its location
for the models we are employing in this paper, and for
simplicity we have adopted unit magnification for all
the stationary phase points. The coordinates of the stationary
phase points can be generated as Normally-distributed
random numbers, using a commercial software package, in
either one or two dimensions. The results, for an assumed
number of $N=100$ contributing stationary phase paths, are shown
in figures 5 and 6 for one- and two-dimensional Gaussian
images, respectively. 

Figures 5 and 6 show a much richer structure than is evident
in their ensemble-average counterparts, figures 1 and 3,
respectively, although the overall power distribution in the
snapshots appears broadly consistent with that in the
ensemble average cases --- as it should, of course. Of particular
interest is that both snapshots manifest inverted parabolic
features; we shall refer to these features as ``arclets''.
The arclets can be understood in the following
way. For the linear image, the $N$ speckles are characterised
by their locations on the $x$-axis: $\theta_i$, $i=1...N$, with
the interference appearing in the secondary spectrum at
$q_{ij}=\theta_i-\theta_j$, $p_{ij}= \theta_i^2-\theta_j^2$. If
we consider a fixed value of $i$ and allow $j$ to vary, then
we recognise that the largest value of $p$ corresponds to
$j=k$ such that $|\theta_k|<|\theta_j|$ for all $j\ne k$, so
that $\theta_k$ lies close to the origin. For this point we
have $q_{ik}\simeq\theta_i$, $p_{ik}\simeq \theta_i^2$, lying
close to the apex of the inverted parabolic arclet. Relative
to this point, we can determine the locations of the other
interference terms, by noting that
\begin{equation}
q_{ij}-q_{ik}=-\theta_j,\qquad p_{ij}-p_{ik}=-\theta_j^2,
\end{equation}
whence
\begin{equation}
p_{ij}-p_{ik}=-(q_{ij}-q_{ik})^2,
\end{equation}
which describes the locus of points in the $i$th arclet as the index
$j$ is allowed to vary. Alternatively, if we vary $i$ and keep $j$
fixed then the locus of points corresponds to
\begin{equation}
p_{ij}+\theta_j^2=(q_{ij}+\theta_j)^2,
\end{equation}
which is an ``upright'' parabola of the same curvature
(up to a sign) as that of the individual arclets. Hence
where there are gaps in the image -- i.e. no stationary
phase points around that value of $\theta$ -- there are
corresponding gaps in the intensity of the arclets, and
those gaps form parabolae cutting through the secondary
spectrum. The particular case where $j=k$ in equation 51
describes a locus very close to that of the apexes of the
arclets: $p_{ik}\simeq \theta_i^2\simeq q_{ik}^2$, so this
locus also exhibits the same curvature, but passes through
the origin of the secondary spectrum ($i=k$). Indeed
the self power for {\it all\/}  image points ($j=i$), appears at
$q=0$, $p=0$, so all of the arclets pass through the origin.
These features are strikingly similar to
what is seen in some of the data (see \S6). 

For the case of a snapshot of an axisymmetric image, the
analysis leading to equation 50 yields instead the locus
\begin{equation}
p_{ij}-p_{ik}=-(q_{ij}-q_{ik})^2 - \theta_{yj}^2.
\end{equation}
For speckles lying close to the $x$-axis we have
$\theta_{yj}^2\ll\theta_{xj}^2$,
and these points all lie close to the inverted parabolic arclet
which defines the upper envelope of power for interference
with the $i$th point: $p_{ij}-p_{ik}=-(q_{ij}-q_{ik})^2$. Referring
to figure 6 we see that there are indeed many points lying below
each arclet, but the arclets themselves remain clearly visible in many
cases. This is because the relative delay of the interfering speckles
is insensitive to the value of $\theta_{yj}$ in the region
$\theta_{yj}^2\ll\theta_{xj}^2$, thus leading to a high density of
power close to the parabolic upper envelope of the arclet. This is the
same effect, described in \S3.3.1, which leads to the visibility of the
single parabolic arc  in figure 4. Similarly, the apexes of
the arclets do not follow a parabolic locus in the case of
an axisymmetric image; instead they are described by
$p_{ik}=q_{ik}^2 + \theta_{yi}^2$, so $p=q^2$ forms a
lower envelope on the possible apex locations.

\subsection{Evolution of features in the snapshot regime}
The arclets shown in figures 5 and 6 are loci of
the interference terms between a given stationary phase point
(speckle in the image) and the line $\theta_y=0$. (More
generally, for a linear image with $\psi\ne0$, the arclets
in figure 5 would correspond to interference with points
close to the locus $\theta_y=\theta_x\tan\psi$.)
In the ``frozen screen'' approximation, the structure of the
phase screen does not evolve, and consequently individual
stationary phase points simply drift across the image as
the line-of-sight moves relative to the screen (Melrose
et al 2004). Correspondingly,
the apex of each of the arclets seen in figures 5 and 6 is
expected to drift across the $(q,p)$ plane. In the case of
an isotropic image the apex of the $i$th arclet drifts along
the parabola $p=q^2 + \theta_{yi}^2$ with $|\dot{q}|\simeq|{\bf v}_\perp|/D_s$
and $\theta_{yi}={\rm const.}$  In the case of a highly
anisotropic image making an angle $\psi=0$ with respect to
the $x$-axis, the motion of the apex of the arclets is along the
parabola  $p=q^2$ with $|\dot{q}|\simeq|{\bf v}_\perp|/D_s$. However, if
$\psi\ne0$ then the misalignment between the linear image
and the velocity vector means that the stationary phase
points cannot remain fixed (even approximately) within
the screen. Instead, each
contributing stationary phase point must slide in a direction
perpendicular to the image so that at every instant these
points constitute a linear image. Thus for highly anisotropic
scattering we expect the apex of each arclet to move along
the parabola $p=q^2\sec^2\psi$, with
$|\dot{q}|\simeq\cos^2\psi\,|{\bf v}_\perp|/D_s$.

Motion of the stationary phase point within the screen, which
occurs in the case of highly anisotropic scattering with $\psi\ne0$,
is of interest because it implies a short lifetime for the
corresponding features in the secondary spectrum (even
in the ``frozen screen'' approximation). The speed of motion
of a given stationary phase point relative to the phase
structure in the screen is $|{\bf v}_\perp\sin\psi|$, in a
direction perpendicular to the image elongation. If the
coherence length measured in this direction is $S_o$
-- which, by definition, is much greater than $s_o$,
the coherence length measured along the direction of
image elongation -- then the corresponding feature in
the secondary spectrum is expected to have a lifetime
of $\sim S_o/|{\bf v}_\perp\sin\psi|$, and this may be
very short unless $\psi$ is very close to $0$.  We note
that in the development in \S3.2, the limit $S_o\rightarrow\infty$
was employed for simplicity. This is, of course, an idealisation
and in practice any $S_o\gg s_o$ suffices to create highly
anisotropic scattering.

The short lifetime of stationary phase
points in highly anisotropic scattering with $\psi\ne0$ suggests
that there may be an observational bias favouring scattering
screens with $\psi\simeq0$, because these are the screens
which yield persistent high-definition arclets. (There may also
be further observational biases favouring $\psi\simeq0$ in the
case of scattering screens of finite spatial extent, because of
the longer interval over which such screens can contribute
to the received radiation.)

The foregoing considerations relate to the evolution of
features in the secondary spectrum under the assumption
of a phase screen which is ``frozen'', i.e. whose phase
profile has no explicit temporal evolution. However, the
frozen approximation may be poor over the time-scale
taken for a given point to move right across the image.
If the phase profile of the screen evolves as a result of
internal motions or wave-modes with velocity dispersion
$\sigma$, then the structure of a coherent patch, of size
$s_o$, is expected to evolve on a timescale $\sim s_o/\sigma$
and this gives us an upper limit on the lifetime of any
feature arising from a single stationary phase point.

\begin{figure}
\includegraphics[width=8cm]{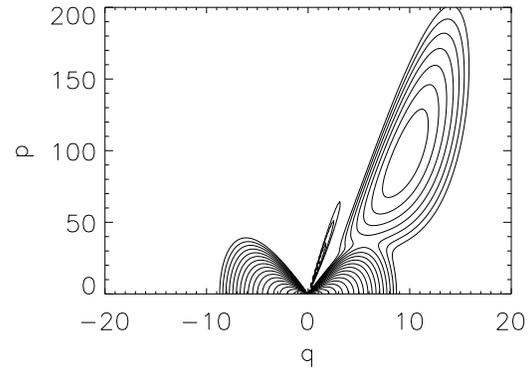}
\caption{Secondary spectrum (delay-Doppler diagram)
for a multiple-component, linear Gaussian image, as described
by equation 53. As in figures 1--6, $q$ is shown here in units
of $\theta_o$, and $p$ in units of $\theta_o^2$.}
\end{figure}

\section{Distinct image components}
To model cases where there are multiple image components, distinct
from the main image centred on ${\bg\theta}=(0,0)$, the anisotropic
image model of \S3.2 is suitable. Here we consider the simplest multiple
component configuration, in which there is only one additional image,
and this component has the same scattering properties (same
characteristic width, $\theta_o$) as the main image; we have chosen
an image with $\psi=0$ and
\begin{equation}
h_1(\theta) = {{99}\over{100}}h(\theta) + {1\over{100}}h(\theta-10\theta_o),
\end{equation}
where $h$ is the Gaussian profile given in eq. 39. In this case the
additional image component is weak (1\% of the main component)
and centred on $\theta_x=10\theta_o$; figure 7 shows the secondary
spectrum for this configuration. Compared with figure 1, the secondary spectrum
for a single-component linear Gaussian image,  there are two new
features in figure 7: the thin linear feature emerging from the origin,
and the elongated blob of power with a peak at $q=10$, $p=100$.
The presence of the latter peak is expected, based on the
analysis presented for speckles in \S4, as it corresponds to the
central portion of the weak component interfering with the central portion
of the main component. However, in contrast to the thin arclets present
in the snapshot example of figure 6, we see that the image represented by
eq. 53 yields a broad swath of extra power extending both to
larger and smaller $p,q$.

We can understand figure 7 in terms of the snapshot properties
(figure 6) by considering the additional image
component to be made up of a large number of speckles, each of
which yields a thin arclet but with the apex at a different location for
each speckle; it then becomes clear that it is the angular extent
of the additional image which causes the power to be smeared out
in the secondary spectrum.  Recall that the angular extent was chosen
to be the same as that of the main image component (eqs. 39, 53),
so the main image and the additional component should exhibit
the same spread in Doppler shift ($q$), but the quadratic mapping
between image coordinates and delay ($p$) means that the additional
image component exhibits a very extended delay spread because
it is not centred on the origin. It is helpful to illustrate this numerically,
considering only the central region of each image component: for the main
image component, extending over $\theta_x=(0\pm1)\theta_o$, the delay
(relative to $(0,0)$) ranges over  $0$ to $1$, whereas for the additional
image component $\theta_x=(10\pm1)\theta_o$ with a delay range of 80 to 120.
This large spread is directly responsible for the large extent in $p$ of the
additional power in figure 7.

The fact that the delay varies linearly across the off-centre image
component means that its self-interference term differs from that
of the main component, even though the two components have the same
shape. At the centre of the additional image component, we have
$\partial p/\partial q=2q=20$, so that the self-interference of this
image component resides in a linear feature sprouting from the
origin of the secondary spectrum, close to the line $p=20q$ .

Although real data do sometimes show an extended swath of power
in the secondary spectrum, spread over a large range of delays, it is
more common to see a number of thin arclets. If these arclets are
transient (see \S4.2) then they may be explicable in terms of
image speckle (\S4), but if not then we need to accommodate
them within the context of distinct multiple image components. We have seen
(\S4) that highly anisotropic images yield thin arclets, because there
is a unique relationship between $p$ and $q$ when we consider
interference with a given speckle. However, when we consider the addition
of an extended, off-centre image component, even in
the limiting case of a linear image, the arclet will exhibit a thickness
(diameter) $\Delta p\sim4q\Delta\theta$, where $\Delta\theta$ is the
angular diameter of the additional image component. The thickness
of the arclet will be greater than this if the anisotropy of the core of
the image is less extreme. The existence of persistent, thin arclets
with apexes at $q\gg1$ would therefore place strong constraints on
the model we have utilised, requiring that the additional image
components introduced to explain the arclets are subject to much
less angular broadening than the core of the image. 

\section{Discussion}
How well do our statistical models (\S3) match the observations? 
For the dozen bright pulsars studied by Stinebring et al (2004), 
scintillation arcs or related phenomena are detected in all of them.
This effectively rules out an axisymmetric Gaussian intensity profile
(Figure~3) since it does not have sufficient power in the wings of the
profile to produce scintillation arcs. (The power distribution near the
origin of the secondary spectrum can be produced by either a Gaussian
or a Kolmogorov profile since they are similar for $\theta \leq \theta_0$.)
A Kolmogorov profile with some degree of anisotropy in the image is a much
better fit to most of the secondary spectra.
For example, in Figure~8  we show a secondary spectrum for a pulsar 
(PSR~B1929+10) that consistently shows well-defined scintillation arcs
with a smooth, symmetric fall-off of power from the origin. This appears
to match best with some form of Kolmogorov image (Figure~2 or 4).
Determining the degree of asymmetry in the image will require a more
extensive comparison of model and data than we attempt here.
A visual comparison of the contour plot with Figure~4, however,
suggests that this image may be close to being axisymmetric.

\begin{figure}
\includegraphics[width=8cm]{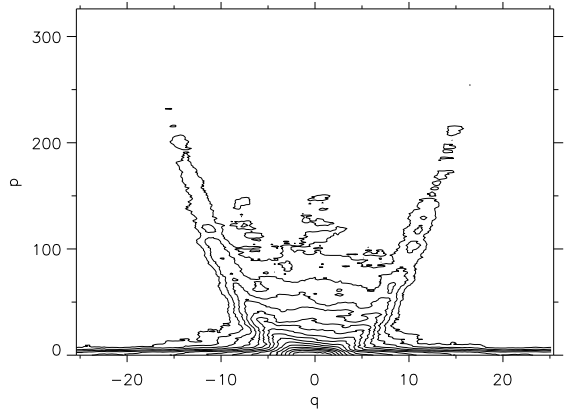}
\caption{The secondary spectrum of PSR~B1929+10 observed at
321~MHz at Arecibo at epoch 2003.65.  Contour levels are spaced
at 3~dB intervals, as for the theoretical results shown in figures 1--4, 7;
the lowest level contour is --45~dB below the maximum.  The observed
frequencies $(f_t,\,f_\nu)$ have been converted to $(q, p)$ (as per
eqns. 28, 29), and scaled to correspond to the axes in figs. 1--7, as
described in the text.
To do so we made use of the measured values $D_p= 0.33 {\rm kpc}$
${\bf v}_p= 163 {\rm km\,s^{-1}}$ for this pulsar, together with the assumptions
that ${\bf v}_p$ dominates the contribution to ${\bf v}_\perp$
and $\beta = 0.37$, and
$\theta_o\simeq0.50\;{\rm mas}$, which are estimated from the
main arc curvature (Stinebring et al 2001) and measured decorrelation bandwidth,
respectively.}
\end{figure}

The most striking secondary spectra are those with pronounced
substructure, particularly those exhibiting inverted arclets. As we
discussed in \S4, these can be reproduced by a model in which
point-like image maxima interfere with broader features near the
origin of the image. In Figure~5, for example, the separation between
isolated image maxima causes gaps, particularly along the delay
($p$) axis, between the inverted arclets. As shown in Figure~9, there
is remarkable qualitative agreement between observations and
the simple point-interference model of \S4. The arclets in Figure~9
are clearly delineated (thin), have apexes that lie along a
$p \propto q^2$ arc that has the same (absolute value of) curvature
as the arclets, and have one side that passes through the origin.
All of these features are consistent with our model. Notice that the
power asymmetry of the arclet (extending further to the right than
to the left) is the mirror image of the power distribution along the
main arc, where the arc has greater power for $-q$ values, but
extends out further on the $+q$ side of the plot.  This is to be expected
if, as in our model, the arclet is the interference between a point-like
feature and the central portion of the image on the sky. The observation
shown here has numerous counterparts for this pulsar over several
years of recent observations, and there is at least one other pulsar
(PSR~B1133+16) that exhibits pronounced arclets at numerous epochs.

\begin{figure}
\includegraphics[width=8cm]{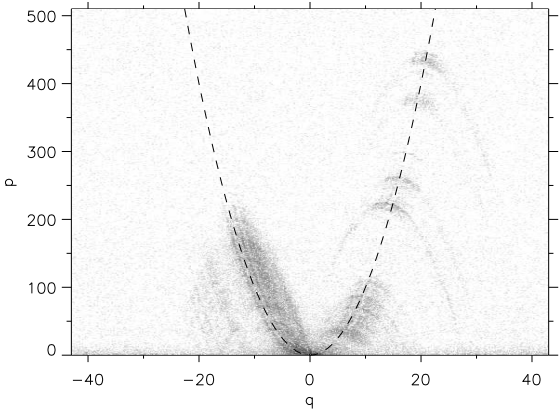}
\caption{The secondary spectrum or PSR~B0834+06 observed at
321~MHz on 2004~Jan~8 with the Arecibo telescope.  The greyscale
is logarithmic in relative power, with white being set 3~dB above the
noise floor and black being set at 5~dB below maximum power.  The
arclet pattern seen prominently on the right hand side of the plot persisted
for more than 25 days during 2004~Jan and moved systematically up and
to the right along the guiding parabolic arc shown by the dashed line. 
The axis scaling is described in the text and can be compared directly
with the plots  of model data. The
axes in this plot are scaled in the same way as in figure 8, but using the
values $D_p= 0.72 {\rm kpc}$, ${\bf v}_p= 175 {\rm km\,s^{-1}}$,
$\beta = 0.33$, and $\theta_o\simeq 0.72\;{\rm mas}$.}
\end{figure}

The models in \S4 were constructed to describe the effects of image
speckle in the snapshot regime, and the evolution of the
corresponding features in the secondary spectrum was discussed
in \S4.2. The timescale on which the arclets in B0834+06 change is
observed to be very long, in some cases at least.  In recent observations
(Hill et al 2004), a persistent arclet pattern has been seen in B0834+06
for more than 25 days.  (This is much longer than either the diffractive
timescale, of 2 minutes, or the refractive timescale, of 2.6 days, for this pulsar
at the observing frequency of 321~MHz.)  Furthermore, the pattern shifts
systematically in the direction from $-q$ to $+q$ in a linear fashion
($|\dot{q}|={\rm constant}$) at a rate consistent with the proper motion
of the pulsar (${\bf v}_\perp\simeq(\beta-1){\bf v}_p$).  These properties
point to a phase profile that is localised in one region of the scattering
screen and is scanned by the pulsar as it moves across the sky; this
circumstance is consistent with the evolution expected in the case
of a ``frozen screen'', as discussed in \S4.2. However, the large
delays associated with the observed arclets (up to $300\;{\rm \mu s}$)
require the phase coherence length to be as small as $s_o\sim10^9\;{\rm cm}$,
if they are caused by diffraction. If we assume that the characteristic
velocity dispersion in the phase screen is $\sigma\ga c_s$, the sound
speed, then with $c_s\sim10\;{\rm km\,s^{-1}}$ we expect that
individual stationary phase points should evolve on a time-scale
$\la10^3\,{\rm s}$. This is 1/2000 of (the lower limit on) their observed
lifetime, and we conclude that the observed features cannot
be attributed to isolated, individual image speckles,\footnote{Strictly
we should be considering isolated {\it pairs\/} of stationary phase points
(Melrose et al 2004), but the essence of the argument remains the same.}despite
the superficial similarity of the data to figure 5. Instead these features
must be attributed to either (i) a collection of many stationary
phase points, if the corresponding image feature has a
diffractive origin, or (ii) one or more stationary phase points
associated with strong refraction. In the former case one can
imagine that the physical cause might be a localised region of
strong turbulence, in which the individual stationary phase points
come and go; the lifetime of the secondary spectrum feature is
then determined by the decay of the turbulence and may be
much longer than that of any individual stationary phase point.
On the other hand, if the feature is associated with strong refraction
-- i.e. a lens -- then the relevant length scale in the phase
screen is $\gg r_F\gg s_o$ and the corresponding timescale
for evolution of the phase screen is naturally much longer. Indeed,
in the case of a lens the phase structure need not be stochastic
and might be in a quasi-steady state with no detectable evolution
(e.g. Walker and Wardle 1998).  In either case the image
component must be tightly concentrated in order to avoid
smearing the power over a range of delays (cf. figure 7).
These compact refracting/scattering structures may be
the same structures which are responsible for the Extreme
Scattering Events (Fiedler et al 1987, 1994).

Despite the lack of agreement with the snapshot model, the results
of \S4 are useful in understanding the inverted arclets.  A comparison
of Figure~9 with Figures~5 and 6 indicates that the image must be
highly elongated in order to produce the observed arclet pattern. 
This is a general result.  No inverted arclet observations exist that
are consistent with an axisymmetric image.  Instead, the presence
of inverted arclets is always accompanied by an absence of power
in the $q=0$, $p>0$ region, and the vertices of the arclets lie along
or very near a definite $p = \eta q^2$ locus, with $\eta$ constant
over many years. All of these aspects follow naturally from an image
that is highly elongated in one dimension and has numerous point-like
features that self-interfere to give the observed arclets.  Furthermore,
the constancy of $\eta$ indicates that the region of the interstellar
medium being probed has a consistent directionality associated with it,
as would be produced by anisotropic scattering in the presence of a
magnetic field (Higdon 1984, 1986; Goldreich \& Sridhar 1995).
This topic is explored further in Hill et al (2004).

One other insight emerges from a comparison between the models
presented here and observations of scintillation arcs (e.g.  Stinebring
et al 2004).  Interference between two identically scattered image
components (e.g. Figure~7) is rare.  Our modeling shows that the
interference between two identically-scattered components would
spread into a broad swath along the scintillation arc due to the wide
range of differential delay values present.  This is rarely seen in
existing data, with thin arclets or pointlike interference features being
much more common.  This important item needs further exploration.

It is evident from the development in \S\S2--4 that secondary spectra
encode information on the structure of radio images (i.e. radio-wave
propagation paths) of pulsars. A great variety of structure is seen in
observed secondary spectra, the vast majority of which remains to be
decoded. Figures 5, 6 and 9 demonstrate that secondary spectra with
thin arclets or other pointlike interference are particularly rich
in information. Referring to equation 15, we see that $N$ points
in the image should yield $N(N-1)/2$ features in the secondary
spectrum. Consequently we have ${\cal O}(N^2)$ constraints but only
${\cal O}(N)$ unknowns, so if $N\gg1$ there is enough information to
deduce the image structure from the observed secondary spectrum.
Each stationary phase point (or point-like structure in the scattering
screen)  yields a measurement of
the phase (up to an overall additive constant), phase gradient, and the
combination of second derivatives which determines the magnification
(eqs. 5--7); all of these quantities being determined at the location
${\bf x}_i$ of the image point. To the extent that the $N$ image points
sample the image plane, these measurements therefore allow us to
deduce the structure of the phase screen, $\phi$.  We emphasise that
this analysis does not require the assumption that the delays are purely
geometric; the large number of available constraints should in fact provide
powerful tests of any approximations which are used in the inversion. The
main obstacle is observational: we require secondary spectra that show
strong arclets or other sharply delineated interference features with good
signal-to-noise ratio throughout. The data shown in figure~9 appear
suitable, but we have not yet attempted the inversion.

\section{Conclusions}
Approximating the Fresnel-Kirchoff Integral by a sum over the
stationary phase points of an arbitrary phase screen has allowed
us to develop a probabalistic description of the power distribution
in pulsar secondary spectra. We have presented a single-integral
formulation of the power distribution for general axisymmetric images,
and a closed form expression for highly anisotropic (i.e. linear) images.
These results are descriptions of the (ensemble-)average properties,
but our development makes it obvious how to generate examples of
snapshot secondary spectra using the Monte Carlo method, and we
have illustrated this technique with two examples. Our results clarify
the origins of the parabolic arcs in observed secondary spectra, and
provide an explanation for the inverted parabolic arclets which are also
sometimes seen.  The observed arclets are due to structured,
highly elongated images. Long-lived arclets require correspondingly
durable features in the phase-screen, and the data show that compact,
isolated clumps of refracting/scattering material must exist in the
interstellar medium. Consideration of the lifetime of highly anisotropic
image components highlights an observational bias in favour of those
which are elongated along the direction of the effective velocity vector.

\section{Acknowledgments}
We have benefitted greatly from helpful discussions with
Barney Rickett and Jim Cordes.

\appendix

\section[]{Evaluation of integrals for axisymmetric images}
In evaluating the integral 
\begin{equation}P=\int d^2{\bg\theta}_1d^2{\bg\theta}_2\,
p({\bg\theta}_1)p({\bg\theta}_2)\,
\delta(p-{\bg\theta}_1^2+{\bg\theta}_2^2)\delta(q-\theta_{1x}+\theta_{2x}),
\end{equation}
one may assume $p\ge0$ without loss of generality, with the
function for $p<0$ determined from that for $p>0$ by $P(-p,q)=P(p,-q)$.
For the axisymetric images under consideration here, the
secondary spectrum is also symmetric in $q$, and in this
Appendix we further
limit the domain of our calculation to $q\ge0$.
One may carry out any two of the four integrals in eq. A.1
over the $\delta$-functions. We evaluate the integrals
in polar coordinates, $d^2{\bg\theta}_i=\theta_id\theta_id\psi_i$.
Carrying out the integral over $d^2{\bg\theta}_1$ gives
\begin{equation}P(p,q)=\sum_\pm\int d^2{\bg\theta}_2\,
\left.
{p({\bg\theta}_1)p({\bg\theta}_2)
\over2|\theta_{1y}|}\right|_{
\theta_{1y}=\pm Y,\,\theta_{1x}=q+\theta_{2x}},
\end{equation}
where $\theta_{1y}$ is determined by the $\delta$-functions,
and is given by
\begin{equation}
Y^2=p+\theta_2^2-(q+\theta_2\cos\psi_2)^2.
\end{equation}
The boundary of the physical region is defined by $Y=0$.
This corresponds to a parabola
with apex at $\theta_{2x}=(p-q^2)/2q$, and asymptotes at
$\theta_{2y}\to\pm\infty$, $\theta_{2x}\to+\infty$. The
integral is over the region outside this parabola. For
$(p-q^2)/2q>0$ the apex of the parabola is to the right
of the origin, and the integral is over $0\le\psi_2<2\pi$
for $\theta_2<(p-q^2)/2q$, and over $\cos\psi_2\le\cos\psi_+$,
where
\begin{equation}
\cos\psi_+={{\sqrt{p+\theta_2^2}-q}\over{\theta_2}},
\end{equation}
for $\theta_2>(p-q^2)/2q$. For $(p-q^2)/2q<0$ the apex of the parabola
is to the left of the origin, so that a region around the origin is
not in the range of integration, and the integral is over
$\cos\psi_2\le\cos\psi_+$ for $\theta_2>|(p-q^2)/2q|$.

One procedure for evaluating the angular integral is
to first change variables from $\psi_2$ to $t$, using
\begin{equation}t=\tan{\psi_2\over2},
\qquad
d\psi_2={2dt\over1+t^2},
\qquad
\cos\psi_2={1-t^2\over1+t^2}.
\end{equation}
Then one finds
\begin{equation}
Y={{(p-q^2+2q\theta_2)^{1/2}}\over{(1+t^2)}}
\left[\bigl(t^2+a^2\bigr)\bigl(t^2+b^2\right)\bigr]^{1/2},
\end{equation}
where
\begin{equation}
a^2={\sqrt{p+\theta_2^2}+q+\theta_2
\over\sqrt{p+\theta_2^2}+q-\theta_2},
\quad
b^2={\sqrt{p+\theta_2^2}-q-\theta_2\over
\sqrt{p+\theta_2^2}-q+\theta_2}.
\end{equation}
For $p>q^2$ and $0<\theta_2<(p-q^2)/2q$ (with $q\ge0$),
we have $a^2>b^2>0$, while for $\theta_2>|(p-q^2)/2q|$
we have $b^2<0<a^2$. Introducing $d^2=-b^2$, and making use
of Gradshteyn and Ryzhik (1965: 3.152, 8.111.2, 8.112.1,
8.113.1) we find that the requisite integrals are
\begin{equation}
\int_0^\infty{dx\over[(a^2+x^2)(x^2+b^2)]^{1/2}}=
{1\over a}\,{\bi K}\left({\sqrt{a^2-b^2}\over a}\right),
\end{equation}
and
\begin{equation}
\int_d^\infty{dx\over[(a^2+x^2)(x^2-d^2)]^{1/2}}=
{1\over\sqrt{a^2-b^2}}\,{\bi K}\left({a\over\sqrt{a^2-b^2}}\right),
\end{equation}
where ${\bi K}$ is the Complete Elliptic Integral, with
\begin{equation}
{\bi K}(k)={\pi\over2}\left(1+{k\over4}+\dots\right).
\end{equation}

The foregoing results imply
\begin{equation}
\int_0^{\pi} {d\psi_2\over Y}=
{1\over{\sqrt{{\cal A}}}}{\bi K}\left(\sqrt{{\cal B}/{\cal A}}\;\right),
\end{equation}
for $0<\theta_2<(p-q^2)/2q$ (with $q\ge0$), and
\begin{equation}
\int_{\psi_{2+}}^{\pi} {d\psi_2\over Y}=
{1\over{\sqrt{{\cal B}}}}{\bi K}\left(\sqrt{{\cal A}/{\cal B}}\;\right),
\end{equation}
for $\theta_2\ge|(p-q^2)/2q|$, where
\begin{equation}
{\cal A}\equiv\left(\sqrt{p+\theta_2^2}+\theta_2\right)^2-q^2
\end{equation}
and
\begin{equation}
{\cal B}\equiv{4\theta_2\sqrt{p+\theta_2^2}}.
\end{equation}
Substituting these
results into equation A2 then yields our final result
for the secondary spectrum corresponding to an
axisymmetric image (equation 43).

\begin{thebibliography}{}

\bibitem[]{} Armstrong~J.W., Rickett~B.J., Spangler~S.R. 1995 ApJ 443, 209 
\bibitem[]{} Cordes~J.M., Rickett~B.J., Stinebring~D.R., Coles~W.A. 2004 (In preparation)
\bibitem[]{} Cordes~J.M., Weisberg~J.M., Boriakoff~V. 1985 ApJ 288, 221
\bibitem[]{} Ewing~M.S., Batchelor~R.A., Friefeld~R.D., Price~R.M., Staelin~D.H. 1970 ApJL 162, L169
\bibitem[]{} Fielder~R.L., Dennison~B., Johnston~K.J., Hewish~A. 1987 Nature 326, 675
\bibitem[]{} Fielder~R.L., Dennison~B., Johnston~K.J., Waltman~E.B., Simon~R.S. 1994 ApJ 430, 581
\bibitem[]{} Goldreich~P., Sridhar~S. 1995 ApJ 438, 763
\bibitem[]{} Goodman~J., Narayan~R. 1989 MNRAS 238, 995
\bibitem[]{} Gradshteyn~I.S., Ryzhik~I.M. 1965 Table of integrals, series and products
\bibitem[]{} Gupta~Y., Rickett~B.J., Lyne~A.G. 1994 MNRAS 269, 1035
\bibitem[]{} Gwinn~C.R., Britton~M.C., Reynolds~J.E., Jauncey~D.L., King~E.A., McCulloch~P.M., Lovell~J.E.J., Preston~R.A. 1998 ApJ 505, 928
\bibitem[]{} Harmon~J.K., Coles~W.A. 1983 270, 748
\bibitem[]{} Hewish~A. 1980 MNRAS 192, 799
\bibitem[]{} Hewish~A., Wolzszczan~A., Graham~D.A. 1985 MNRAS 213, 167
\bibitem[]{} Higdon~J.C. 1984 ApJ 285, 109
\bibitem[]{} Higdon~J.C. 1986 ApJ 309, 342
\bibitem[]{} Hill~A.S., Stinebring~D.R., Barnor~H.A., Berwick~D.E., Webber~A. 2003 ApJ 599, 457
\bibitem[]{} Hill~A.S. et al 2004 (In preparation)
\bibitem[]{} Lee~L.C., Jokipii~J.R. 1975 ApJ 201, 532
\bibitem[]{} Lee~L.C., Jokipii~J.R. 1976 ApJ 206, 735
\bibitem[]{} Melrose~D.B, Macquart~J.-P., Zhang~C.M., Walker~M.A. 2004 MNRAS (Submitted)
\bibitem[]{} Narayan~R., Hubbard~W.B. 1988 ApJ 325, 503
\bibitem[]{} Rickett~B.J. 1990 ARAA 28, 561
\bibitem[]{} Rickett, B.J., Lyne~A.G., Gupta~Y. 1997 MNRAS 287, 739
\bibitem[]{} Roberts~J.A., Ables~J.G. 1982 MNRAS 201, 1119
\bibitem[]{} Shishov~V.I. 1974 Sov Astron AJ 22, 544
\bibitem[]{} Stinebring~D.R., McLaughlin~M.A., Cordes~J.M., Becker~K.M., Espinoza Goodman~J.E., Kramer~M.A., Sheckard~J.L., Smith~C.T. 2001 ApJL 549, L97
\bibitem[]{} Stinebring~D.R. et al 2004 (In preparation)
\bibitem[]{} Walker~M.A., Wardle~M.J. 1998 ApJL 498, L125
\bibitem[]{} Wolszczan~A., Cordes~J.M. 1987 ApJL 320, L35

\end{thebibliography}
\end{document}